\documentstyle[aps,prb,epsf]{revtex}

\begin{document}
\twocolumn[\hsize\textwidth\columnwidth\hsize\csname
@twocolumnfalse\endcsname
                                                  
\title{Longitudinal and transverse dissipation in a simple model for the vortex lattice
with screening}
\author{E. A. Jagla}
\address{Centro At\'omico Bariloche and Instituto Balseiro \\
Comisi\'on Nacional de Energ\'{\i}a At\'omica\\
(8400) S. C. de Bariloche, R\'{\i}o Negro, Argentina}
\maketitle

\begin{abstract}
Transport properties of the vortex lattice in high temperature superconductors 
are studied using numerical simulations in the case in which the non-local interactions 
between vortex lines are dismissed. The results obtained for the longitudinal 
and transverse resistivities in the presence of quenched 
disorder are compared with the results of experimental measurements and 
other numerical simulations 
where the full interaction is considered. This work shows that the 
dependence on temperature of the resistivities is well described by the 
model without interactions, 
thus indicating that many of the transport characteristics of the vortex structure 
in real materials are mainly a consequence of the topological configuration of the 
vortex structure only. In addition, for highly anisotropic samples, a regime is
obtained where longitudinal coherence is lost at temperatures where transverse 
coherence is still finite. I discuss the possibility of observing this regime 
in real samples.
\end{abstract}

\vskip2pc] \narrowtext

\section{Introduction}

Physical properties of the vortex structure in high temperature
superconductors are strongly dependent on the values of some external
parameters, such as magnetic field, anisotropy, and quenched disorder.\cite
{blatter} Some of these properties -as for example the existence of a first
order melting transition in clean samples-\cite{1orden} are originated in
the interactions between vortices. Some others, however, are rather
independent of the details of the interaction, and are related to the
topological configuration of the vortex structure. It is interesting then to
analyze in the simplest way the origin of those properties that depend only
on the geometrical configuration of vortices, and not on their interaction.
I concentrate in this paper on the behavior of the linear resistivity $\rho$ 
($\rho=\lim_{I\rightarrow 0} V/I$) as a
function of temperature -which has been widely used experimentally to
unravel the properties of the vortex structure-,\cite{blatter} both
perpendicular and parallel to the applied magnetic field of a model that
disregards non-local interactions between vortices. The results of numerical
simulations on this model compare well with
experimental results obtained in Y$_1$Ba$_2$Cu$_3$O$_7$ (YBCO, rather low
anisotropy) and Bi$_2$Sr$_2$Ca$_1$Cu$_2$O$_8$ (BSCCO, high anisotropy)
samples, as long as we consider zones of the phase diagram of the materials
in which a first order transition is not observed. As stated above, the
first order transition is generated by the interaction of vortices, and
cannot be expected to occur in a model with only local interactions.

I describe the model in the next section, and the results in section III.
The case of very high anisotropies deserves special attention and the
limit of two-dimensional systems is discussed in section IV. The relevance
of these results to real materials (in which vortices interact at finite
-usually large- distances) is discussed in section V. Finally, in section
VI, I summarize and conclude.

\section{Model}

Vortices are modeled at different levels of detail when performing numerical
simulations. A quite precise description is the Ginzburg-Landau theory,
formulated in terms of the superconducting order parameter $\Psi \equiv
\left| \Psi \right| \exp \left( i\theta \right) $.\cite{gl} In this context,
when an external magnetic field or thermal fluctuations are introduced,
vortices appear in the system as line-singularities around which $\oint
\theta \left( {\bf r}\right) d{\bf r}=2\pi $. A usual simplification which
is appropriate in high temperature superconductors is to consider the
modulus $\left| \Psi \right| $ of the order parameter as a constant, and
keep only the phases $\theta \left( {\bf r}\right) $ as the dynamic
variables. This leads to the study of the uniformly frustrated $XY$ model,%
\cite{xydegl} which has been extensively studied, both because of its
intrinsic properties and because of its applications to superconducting
systems. As first shown by Villain,\cite{villain} the most important degrees
of freedom of this model can be identified with the positions of the vortex
in the system and an alternative description with these positions as the
fundamental dynamical variables can be obtained. The original structure of
the Ginzburg-Landau free energy is reflected at this level in the particular
form of the interactions between vortices,\cite{carneiro,korshunov} which are cut off
at distances of the order of the some penetration lengths $\lambda _{ab}$, $%
\lambda _c$ (the subindexes refer to the crystalline directions). These
distances are in general large compared with the intervortex distance, and
so the energy of the system has contributions coming from interaction
between vortices that are far away from each other. I will consider here the
case in which $\lambda $ is very small or, stated in another form, when the
non-local terms of the interaction are dropped. In this case, the energy of
a given configuration is simply proportional to the total length of vortices
in the system. The investigation of the properties of the model in this case
is important since, as we shall see, it gives insight on the behavior of the
system with the full interaction, and allows to understand that the origin
of many properties of the vortex lattice comes from the topological structure
of vortex lines, and not from the exact nature of the interactions.

I will consider vortex segments lying on the bonds of a cubic mesh with
periodic boundary conditions. Formally, the Hamiltonian of the system is

\begin{equation}
H_0=\sum_{i,\mu }{}^{\prime }\varepsilon _{i,\mu }\left( n_{i,\mu }\right)
^2,  \label{hamil}
\end{equation}
where $n_{i,\mu }$ are integer variables defined on the nodes $i$ of the
lattice, with direction $\mu $ ($\mu =$ {\it a, b, c}). The prime in the sum
symbol indicates that only those configurations with zero divergence of the
vector field $n$ have to be considered. The constants $\varepsilon _{i,\mu }$
are the energies of vortex segments at the positions $i,\mu $. We allow for
the existence of anisotropy, defining a parameter $\eta $ as $\eta \equiv
\left\langle \varepsilon _{i,c}\right\rangle /\left\langle \varepsilon
_{i,ab}\right\rangle ,$ with $\left\langle ...\right\rangle $ indicating
averaged values throughout the lattice. Disorder is introduced by allowing
the values of $\varepsilon _{i,\mu }$ to be different in different points of
the sample, fluctuating around the mean value. A disorder parameter (the
same for the three spatial directions) is defined as $D=\left( \varepsilon
_{i,\mu }^{\max }-\varepsilon _{i,\mu }^{\min }\right) /\left( \varepsilon
_{i,\mu }^{\max }+\varepsilon _{i,\mu }^{\min }\right) ,$ with $\varepsilon
_{i,\mu }^{\max }$ and $\varepsilon _{i,\mu }^{\min }$ being the maximum and
minimum value of the energy of a vortex segment. The distribution between $%
\varepsilon _{i,\mu }^{\min }$ and $\varepsilon _{i,\mu }^{\max }$ is taken
flat.

As the initial configuration of the system, a set of straight vortex lines
directed along the {\it c} direction and uniformly distributed on the {\it ab%
} plane is considered. The number of vortices divided by the number of elementary
plaquettes of the system perpendicular to the
{\it c} direction defines the dimensionless magnetic field $H$. 
The Montecarlo process for updating the configuration
consists in sequentially proposing the creation of elemental squared loops
in all plaquettes of the lattice and with the three possible directions. The
acceptance of the new configuration is carried out using a standard
Metropolis algorithm. The initial configuration and the Montecarlo procedure
guarantee that at any moment the vortex configuration has zero divergence.
In order to calculate resistivities, both parallel and perpendicular to the
applied field ($\rho _c$ and $\rho _{ab}$, respectively), we have to include
a small external current $I$. This is done by adding a term to the
Hamiltonian (\ref{hamil}) that changes the energy of loops oriented
perpendicularly to the current. One orientation increase its energy by $+I,$
and the other one decreases it in the same quantity. The value of the
external current $I$ is chosen in such a way that the disbalance between
right- and left-handed loops is never higher that 1/100 of the energy of the
loop. In this regime of low currents the response is linear in the applied
current, and resistivities do not depend on the exact value of $I$. The
numerical results for different values of the anisotropy follow.

\section{Results for three dimensional samples}

I will describe the results of the numerical simulations performed in
systems with progressively higher anisotropies. Three different regions are
distinguishable.

\begin{figure}
\narrowtext
\epsfxsize=3.3truein
\vbox{\hskip 0.05truein
\epsffile{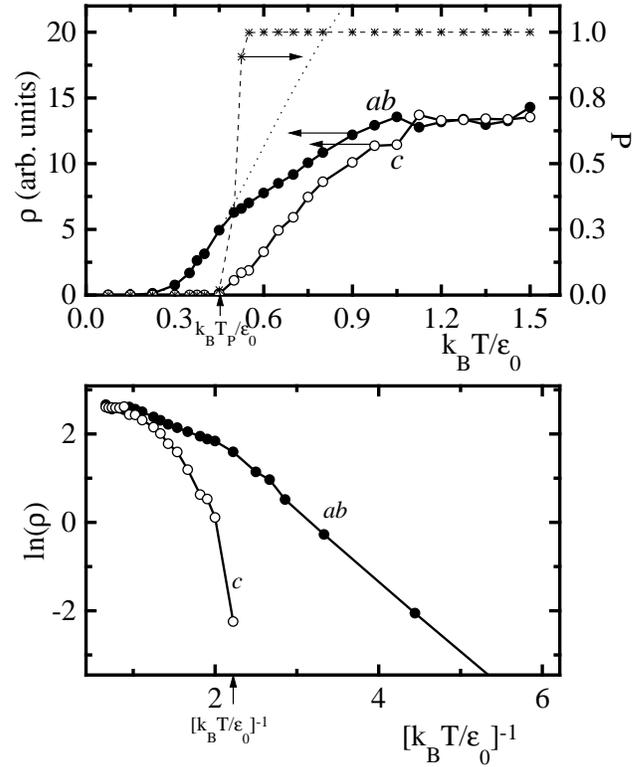}}
\medskip
\caption{(a) Circles: resistivities along the {\it ab} and {\it c}
directions using the Hamiltonian (\ref{hamil}) for an isotropic lattice of
size $40\times 40\times 8$ with $D=0$ and $H=1/4$. The dotted line is a
fitting to a thermally activated movement of single vortices of the form $%
\rho _{ab}\sim \frac 1T\exp \left( -2\varepsilon _0/k_BT\right) $. Stars:
percolation probability as defined in text, note that this magnitude is
(different from) zero only if $\rho _c$ is (different from) zero. (b) The
resistivity values plotted to show thermally activated behaviors.}
\label{rabrcdet}
\end{figure}

For low anisotropies, typical results for $\rho _c$ and $\rho _{ab}$ as a
function of temperature are shown in Fig. \ref{rabrcdet} for a sample of
size $L_a\times L_b\times L_c=40\times 40\times 8$ with $D=0$ (all $%
\varepsilon _{i,\mu }$ equal to $\varepsilon _0$) and $H=1/4$. As we see the 
{\it ab} dissipation has a thermally activated behavior (dotted line
fitting) as long as the {\it c} axis dissipation is zero ($T<T_P$). The
activation energy of this process is $2\varepsilon _0,$ and corresponds to
the energy necessary to create a double kink on a straight vortex, which is
the first step in the process of movement. So in this region the dynamics of
the system is that of a set of individual vortices thermally wandering into
the sample. At a certain temperature $T_P,$ the {\it c} axis dissipation $%
\rho _c$ becomes finite. The origin of this dissipation is related to a
percolation transition of vortices occurring in the system, as studied
before in the three dimensional Josephson junction array model.\cite{perco}
For $T<T_P$ the vortices are practically independent, and a {\it c} axis
directed current exerts no net force on them, thus generating no dissipation
(in the linear regime). For $T>T_P$ the vortex lines are so heavily
interconnected that vortex paths running (on average) in the {\it ab}
direction appear. An external current applied along the {\it c} direction,
being perpendicular to these paths, exert a net force on them and generates
dissipation. In Fig. \ref{rabrcdet}(a) we also see the value of the
percolation probability $P$, defined as the fraction of time in which at
least one percolation path is found in the system. It is clearly observed
that the {\it c} axis dissipation is zero if $P$ is zero. The transition in
the $P\left( T\right) $ curve between 0 and 1 becomes sharp in the limit $%
L_{ab}\rightarrow \infty $.\cite{perco}

An interesting effect of the percolation transition on the values of $\rho
_{ab}$ is observed. When $\rho _c$ starts to be different from zero, the
values of $\rho _{ab}$ deviate from the prediction of an activated behavior
of individual vortices, and become smaller. This is an indication that for
temperatures greater that $T_P$ the percolation of the vortex structure and
the existence of many thermal excitations interferes with the thermally
activated behavior of single vortices. In fact, for $T>T_P$ the concept of
an isolated vortex in the system looses its sense, because we have a
strongly entangled configuration of vortex lines. This effect of the
percolation transition on the $\rho _{ab}\left( T\right) $ curve causes this
to develop a typical shoulder that has been experimentally observed in
measurements on YBCO samples,\cite{hombroexp,fm} although its origin had not
been clearly established.

In the case $D=0,$ when increasing anisotropy, the {\it ab} plane activation
energy goes to zero as $1/\eta $ because this is mainly determined by the
energy necessary to create a double kink, which is precisely $2/\eta $. Also
the {\it c} axis transition is governed by an energy scale of the order of $%
\sim 1/\eta ,$ because this temperature is mainly determined by the typical
energy of the interlayer excitations (which goes as $\sim 1/\eta $)$.$ On
heating, and from a practical point of view, the {\it c} axis transition
occurs at a temperature at which $\rho _{ab}$ is clearly different from zero
for all anisotropies.

The physics is richer in the case of samples with defects ($D\neq 0$). In
this case, when anisotropy is increased, the activation energy for the {\it %
ab} dissipation tends to a value of the order of $D^{1/2},$ because the
energy of a vortex segment piercing the {\it ab} plane is not constant in
this case but has a dispersion of the order of $D^{1/2},$ and this is the
typical energy barrier that has to be overcome when a vortex segment wanders
within the {\it ab} planes. Anisotropy decreases the percolation temperature 
$T_P$ in a factor $\sim 1/\eta $, but has minor effect on the thermal
activation in the planes as long as $D\neq 0$. Thus for high enough
anisotropies, the {\it c} axis dissipation is expected to occur even at
lower temperatures than the {\it ab} plane dissipation. However, when this
range is approached a particular transition occurs in the system. The low
temperature configuration of the vortex structure passes from a
disentangled, and rather ordered configuration of vortex lines for low
anisotropies, to an entangled configuration of vortex lines, i.e., we can
say that the percolation temperature of the system drops abruptly to zero.
The origin of this particular transition is the following. If anisotropy is
low the system will prefer to remain ordered, with vortices almost straight
in order to minimize their line energy. If anisotropy is increased,
entangled configurations (in which vortices use the strongest pinning sites
on the {\it ab} planes) diminish their energy and become locally stable, and
from some value of the anisotropy, one entangled configuration becomes
globally stable. This is the critical anisotropy $\eta _1$. This transition
is related to the Bragg glass to vortex glass transition\cite{bragglass}
proposed to occur in very anisotropic samples when increasing the magnetic
field, because -as it has been discussed elsewhere-\cite{jb5} an
increase in the magnetic field is equivalent to an increase in the effective
anisotropy and disorder of the system.

\begin{figure}
\narrowtext
\epsfxsize=3.3truein
\vbox{\hskip 0.05truein
\epsffile{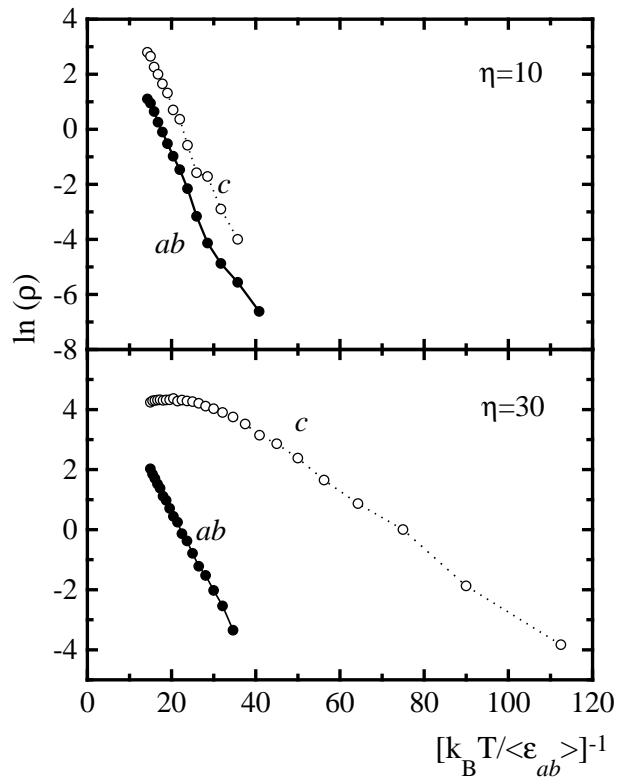}}
\medskip
\caption{Typical results for $\rho _c$ and $\rho _{ab}$ as a function of
temperature for two different values of anisotropy, in a sample of size $%
40\times 40\times 8$, with $D=0.5$ and $H=1/4$.}
\label{rabrcamedio}
\end{figure}

Numerical simulations indicate that in the $\eta >\eta _1$ regime {\em both} 
{\it ab} plane and {\it c} axis dissipation have a thermally activated
behavior as shown in Fig. \ref{rabrcamedio}. The results for activation
energies of $\rho _c$ and $\rho _{ab}$ as a function of anisotropy for a
system of $40\times 40\times 8,$ with $D=0.5$ are shown in figure \ref{eact}
(results for other values of $H$ are similar, with only a rescaling of the 
anisotropy axis). From a numerical (or experimental) point of view, it must 
be kept in mind
that a thermally activated behavior with a given activation energy can be
checked only for values of $k_BT$ greater than some fraction (which depends
on sensibility) of the activation energy. In particular, no statements can
be made about the possible existence of a true critical temperature much
lower than that corresponding to the activation energy. In Fig. \ref{eact}
and for $\eta <\eta _1\sim 4$ the {\it c} axis activation energy is not
defined, because the transition is a percolation process and not a thermally
activated process, as discussed before.

\begin{figure}
\narrowtext
\epsfxsize=3.3truein
\vbox{\hskip 0.05truein
\epsffile{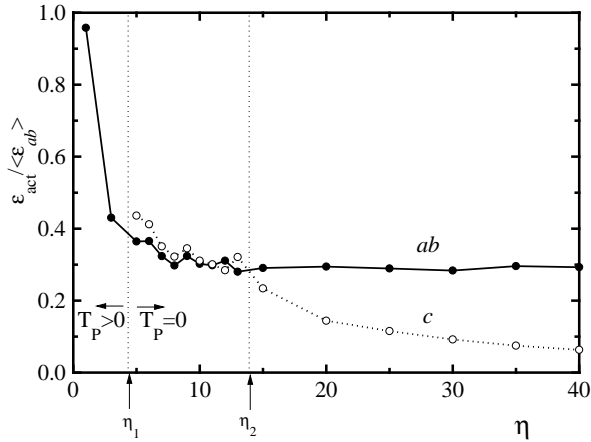}}
\medskip
\caption{Activation energies for the dissipation along {\it c} and {\it ab}
direction as a function of anisotropy (size $40\times 40\times 8$, $D=0.5$, $%
H=1/4$). For $\eta <\eta _1$ the {\it c} axis dissipation is not thermally
activated but has a critical behavior instead. For $\eta >\eta _2$ the {\it c%
} axis activation energy decays as $1/\eta $.}
\label{eact}
\end{figure}

The most direct and important conclusion from this figure is that there are
two sub-regions in the range $\eta >\eta _1$. For $\eta _1<\eta <\eta _2$
activation energies for transport parallel and perpendicular to the field are
rather similar, whereas for $\eta >\eta _2$ the {\it c} axis activation
energy is lower than the corresponding to the {\it ab} plane. In the whole
range $\eta >\eta _1,$ {\it ab} plane activation energy is rather anisotropy
independent, indicating that the {\it ab} plane dissipation is governed by
the thermal activation of vertical segments of vortex lines pinned to the 
{\it ab} planes. For $\eta <\eta _2$ both parallel and perpendicular
dissipation have the same activation energy indicating that they both
originate in the same physical process.\cite{koshelev} In fact, {\it ab}
plane dissipation is caused by the thermal depinning of vortex lines that
cross the sample in the {\it c} direction. In turn, the {\it c }axis
dissipation is caused by the thermal depinning of vortex lines that cross
the sample in the {\it ab} direction (which exist because the vortex
configuration is heavily entangled). However, as the formation of these
paths is mediated by the externally generated vortices they are also pinned
to the {\it ab} planes, and the activation energy for both processes is the
same. Despite this, a global factor in the resistivity relation $\rho
_c/\rho _{ab}$ depending upon the anisotropy and the geometrical
configuration of vortex lines (especially on the relation between number of
paths running along {\it c} axis and {\it ab} direction) is expected.

For $\eta >\eta _2$ the decrease of the activation energy for {\it c} axis
dissipation as $1/\eta $ indicates that a new dissipation mechanism that
depends only on the excitation of horizontal loops between planes is taking
place. Being the vertical segments of vortex lines still frozen in the range
of temperatures in which the {\it c} axis dissipation starts to be
appreciable, this mechanism has to be related to processes occurring between
consecutive {\it ab} planes, {\it i.e.}, in the zone $\eta >\eta _2$ a
complete {\it decoupling} of the planes takes place.

\section{Perpendicular dissipation in two dimensional systems}

As I mentioned in the previous section, the activation energy of the {\it c}
axis dissipation at high anisotropies ($\eta >\eta _2$) goes to zero as $%
1/\eta ,$ indicating that processes involving only the horizontal segments
between consecutive planes are important. In order to understand clearly
this kind of processes I will analyze the dissipation in the case of a
unique horizontal plane.

Consider the model studied previously (Eq. \ref{hamil}) but now on a two
dimensional geometry. It is useful to point out that this model (in the case
without disorder) has been named in other context the {\it roughening model} of
surfaces.\cite{rough,ng,rough0} The mapping is made between vortex segments
in the original model and height differences of a growing surface in the
roughening model, the divergence free condition in the vortex model being
the key property that makes possible the mapping. In the language of the
roughening problem, the surface is smooth at large scales in the low
temperature phase, and is rough in the high temperature phase. The long
extending terraces that make the surface rough are no more than the infinite
length vortex paths of the vortex model, and the growing rate of the surface
maps onto the perpendicular resistivity of the vortex model. In the
homogeneous case (without disorder), the roughening model is known to have
an inverted Kosterlitz-Thouless transition at a temperature $T_{KT}\simeq $ $%
0.36\epsilon $ where $\epsilon $ is the energy of an elemental growing step
(an elemental vortex loop in the vortex model, $\epsilon =4\varepsilon _0$).%
\cite{rough1} This temperature can be viewed again as a percolation
temperature of vortex lines in the vortex system: below the critical
temperature $T_{KT}$ there is no infinite length paths, whereas for $T>T_{KT}
$ these paths exist. In Fig. \ref{2dd0} the percolation probability for
systems of different sizes is shown, and the transition is clearly
distinguishable. As in the three dimensional case the transition in the
variable $P\left( T\right) $ becomes sharp in the limit $L_{ab}\rightarrow
\infty $. From known results on the roughening model\cite{ng} we can
directly conclude that for an infinite system the perpendicular 
resistivity ($\rho _c$) jumps from zero to a finite value at the temperature 
$T_{KT}$.

\begin{figure}
\narrowtext
\epsfxsize=3.3truein
\vbox{\hskip 0.05truein
\epsffile{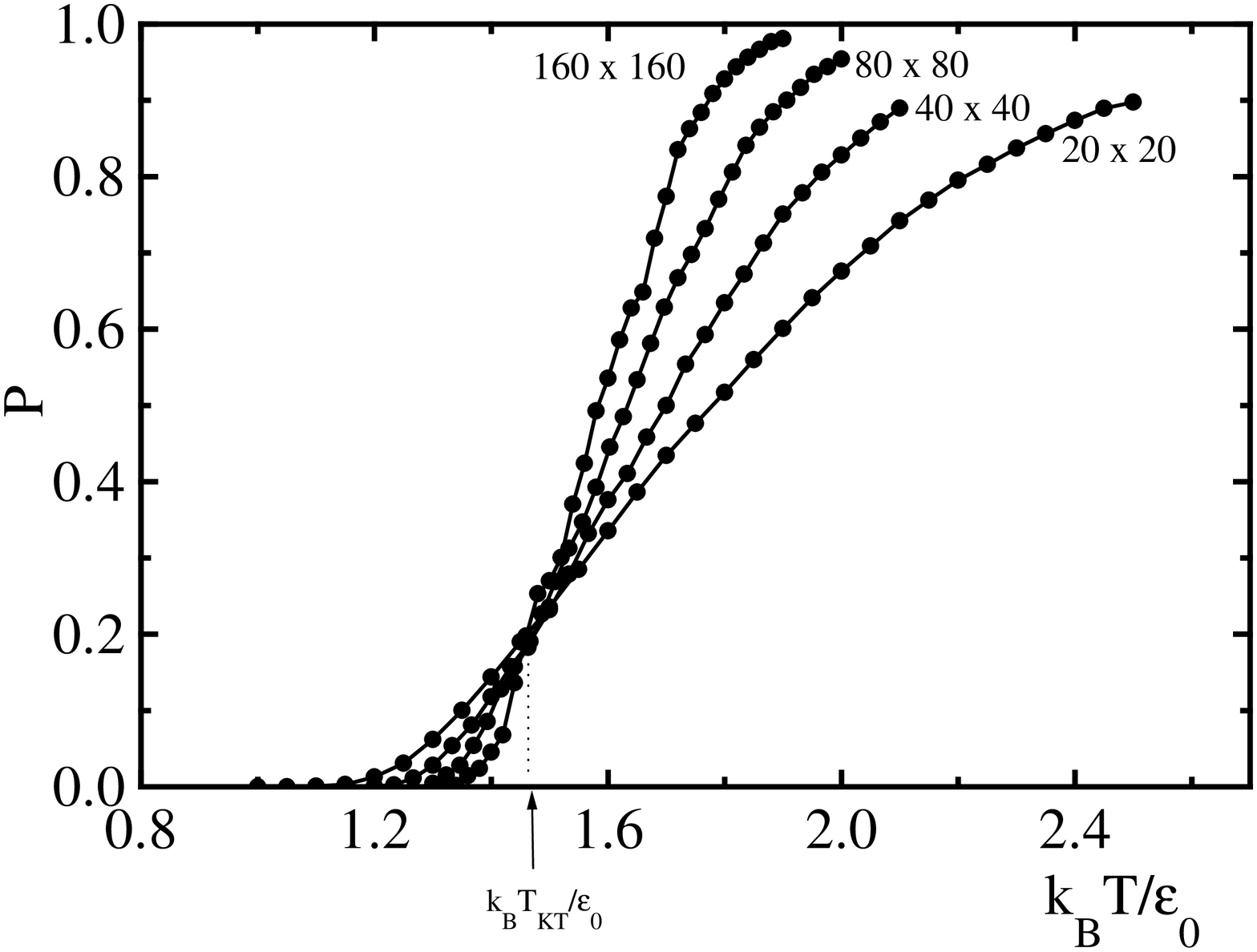}}
\medskip
\caption{Percolation probability in a two-dimensional system without
disorder for different system sizes. A phase transition in the limit of very
large systems is clearly observed.}
\label{2dd0}
\end{figure}

The disorder in the two-dimensional case may have two origins: one is the
disorder introduced directly in the Hamiltonian, through the dependence of $%
\varepsilon _{i,\mu }$ on coordinates (Eq. (\ref{hamil})). The other comes
from the existence in the 2D system of some quenched horizontal segments
that come from the horizontal parts of the 3D vortices. This segments within
the 2D system{\it \ }do not satisfy the condition $\nabla n=0,$ so its
inclusion in the dynamics makes a non trivial contribution. We will
concentrate on the second type of disorder for two reasons: numerical
simulations show that the qualitative changes produced by the two types of
disorder are similar, and secondly because this type of disorder is
dynamical, and may change when changing anisotropy.

So the problem may be stated as that of a Hamiltonian 
\begin{equation}
H=\sum_{i,\mu }{}^{\prime }\varepsilon _0\left( n_{i,\mu }-b_{i,\mu }\right)
^2,  \label{h2}
\end{equation}
where now $b_{i,\mu }$ represent the horizontal segments induced by the 3D
vortices, and $\mu =a,b$. Note that the variables $n_{i,\mu }$ still satisfy
the condition $\nabla n=0$, and the bare energy of the segments lying on the
links has been taken to be $\varepsilon _0$ in all sites. It is interesting
to consider the behavior of the system when the number of horizontal
segments is being increased. We can characterize this value as the fraction
of links in which $b_{i,\mu }=\pm 1,$ (which we denote by $D$). A necessary
condition for a finite dissipation (perpendicular to the plane) is still
that the vector field $n_{i,\mu }$ generates a path running all across the
(two-dimensional) system. The probability of existence of such paths is
shown for different sample sizes and different values of the disorder in
Fig. \ref{pcded}(a). For $D < D_{cr} \simeq 0.3$ a well defined transition when the
system size increases is detected. For finite disorder the transition
temperature $T_P$ decreases with respect to the zero disorder value $T_P^{(D=0)}=T_{KT}\simeq
1.47$. If disorder is too high, however, there is no intersection of lines
corresponding to different sizes, indicating that the percolation transition 
moves down to
aero temperature. The behavior of the percolation temperature as a
function of $D$ is plotted in Fig. \ref{pcded}(b).

\begin{figure}
\narrowtext
\epsfxsize=3.3truein
\vbox{\hskip 0.05truein
\epsffile{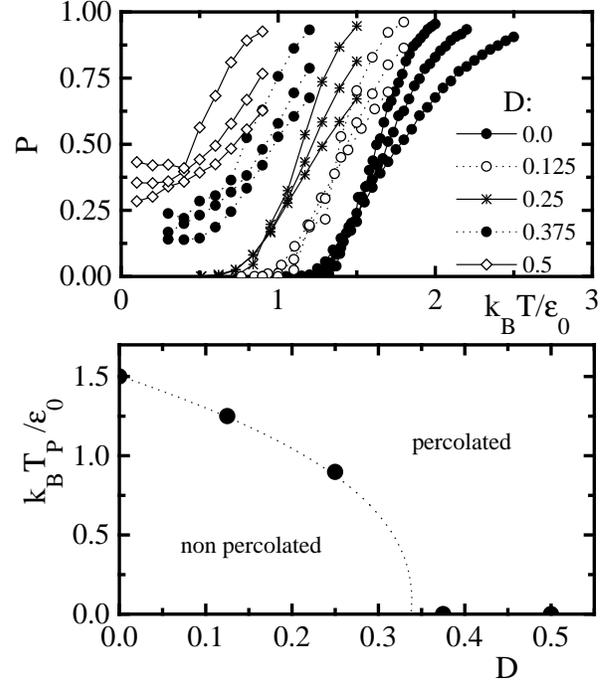}}
\medskip
\caption{(a) Percolation probability in a two-dimensional system for
different values of the disorder for system sizes $20\times 20$, $40\times 40
$, and $80\times 80$ (from bottom to top, at the right of the curves). 
The results are the averaged values over 3
configurations of disorder for $D=0.125,$ $0.25,$ and over 8 configurations
for $D=0.375,$ $0.5$. For $D<D_{cr}\simeq 0.3$ a percolation phase
transition at finite temperature is clearly detectable. For higher values of
disorder percolation temperature moves down to $T=0$. (b) Temperature of the
percolation transition as a function of disorder as obtained from the
results in panel (a). Points are the results of numerical simulations.
Dotted line is a guide to the eye only.}
\label{pcded}
\end{figure}

Perpendicular resistivity simulations in this model (Fig. \ref{r2d}) give
values that clearly go to zero in the case in which $T_P$ is finite ($D<0.3$)\cite{otranota}, 
and in the case when this variable is finite at any
temperature give results that can be well fitted by a thermal activation
expression. However, the
possibility that the system has a real critical temperature at a temperature
where the resistivity becomes undetectable in the simulations cannot be
ruled out (see the note on the mapping to the roughening problem with
disorder in the next paragraph).

\begin{figure}
\narrowtext
\epsfxsize=3.3truein
\vbox{\hskip 0.05truein
\epsffile{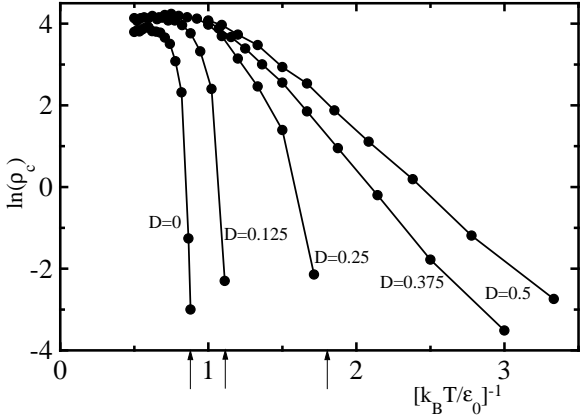}}
\medskip
\caption{Perpendicular resistivity of a two dimensional system ($80\times 80$%
) as a function of temperature for different values of the concentration of
quenched segments $D$. For $D<D_{cr}\simeq 0.3$ the approximate temperature at which
temperature vanishes is indicated by arrows. For $D>D_{cr}$ the curves show activated 
low temperature tails.}
\label{r2d}
\end{figure}

The behavior of the percolation probability and resistivity for different
values of $D$ of model (\ref{h2}) can be qualitatively explained using an
argument of the type of the celebrated original Kosterlitz-Thouless argument
for the phase transition in the two-dimensional $XY$ model. It goes as
follows. The energy of a path running across a system of size $\sim L$ is
the sum of $\sim L$ variables. According to (\ref{h2}), each of these variables may have the value $%
\varepsilon _0$, $-\varepsilon _0$, or $3\varepsilon _0$ depending if for
that particular size the variable $b$ is $0,\pm 1$. The energy of the path
becomes a Gaussian variable with mean value $\sim L$ and dispersion $\sqrt{%
LD\varepsilon _0}$. The total number of paths of length $\sim L$ running
across the system is exponential in $L$, i.e., about $\sim e^L$. Taking into
account these estimations, and considering the calculated number of paths
with different energies as a density of states for non-interacting
``particles'', it is possible to make the statistical mechanics of the
system and look for the thermodynamic free energy of the equilibrium
configuration. I only show the results, and not the details of the
calculations, which are straightforward. Two clearly different regions
appear (see Fig. \ref{ktargumento}): there is a zone (labelled I) of high
disorder or high temperature in which some percolation paths have negative
free energy, and so their existence is thermodynamically stable. In the other
zone (labelled II) with low temperature and disorder, all percolation paths
have positive free energy and thus the configuration without any path is the
stable one. In zone II the system does not contain any percolation paths,
and its resistivity is zero. When passing to zone I increasing the
temperature, many percolation paths become immediately accessible at the
transition, and perpendicular resistivity jumps to a finite value. In the
case $D>D_{cr}$ there are available percolation paths down to zero
temperature, and its likely that resistivity remains finite (with a
thermally activated behavior) up to zero temperature, although a critical
behavior of the resistivity at some finite temperature cannot be completely
ruled out from these qualitative arguments.\cite{notita} These estimates
compare well with the numerical results on Hamiltonian (\ref{h2}) (Figures (%
\ref{pcded}) and (\ref{r2d})).

\begin{figure}
\narrowtext
\epsfxsize=3.3truein
\vbox{\hskip 0.05truein
\epsffile{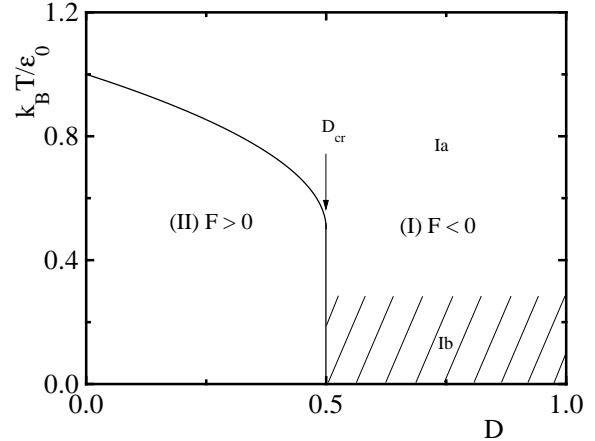}}
\medskip
\caption{Qualitative phase diagram from the estimation of free energy of
percolation paths, as discussed in the text. Zone I is a region with finite
linear dissipation. Zone II has zero linear dissipation. The labels Ia and
Ib and the dashed region are linked to the discussion presented in [18].}
\label{ktargumento}
\end{figure}

The mechanism discussed in this section for the perpendicular dissipation of
a two-dimensional system is the responsible of the c axis dissipation in the
three dimensional case for $\eta >\eta _2$ discussed in the previous
section. As it was indicated before this mechanism relies on the movement of
vortex segments between consecutive {\it ab} planes, contrary to the dissipation 
occurring for $\eta < \eta_2$ which involves paths that occupy a finite thickness
along the {\it c} direction.

\section{Relevance to real materials. Effect of non local interactions}

In spite of its simplicity, the previously studied model reproduces many
characteristics of real transport measurements in high-$T_c$
superconductors, indicating that some of these properties are rather
independent of the exact form of the interaction, and they are only
consequences of geometrical properties. In particular, the case $\eta <\eta
_1$ (Fig. \ref{eact}) gives results which are comparable to experimental results in YBCO.\cite
{fm,exper1} In this case the {\it c} axis resistivity has been observed to
appear only at higher temperatures than the {\it ab }dissipation, and a
shoulder in $\rho _{ab}$ near $T_P$ similar to that of Fig.\ref{rabrcdet} has been
experimentally observed. Numerical simulations using the complete
vortex-vortex interaction give also the same qualitative behavior.\cite
{numer1,perco} The results of the previous sections for the activation energy in
the range $\eta _1<\eta <\eta _2$, namely the coincidence of activation
energies for $\rho _{ab}$ and $\rho _c,$ has been experimentally observed in
BSCCO samples\cite{exper2} and theoretically interpreted as an indication of
the same origin for both dissipations in this regime.\cite{koshelev}

The only one region that gives a qualitatively new result is the range $\eta
>\eta _2$, where a finite value for $\rho _c$ was obtained even at
temperatures at which $\rho _{ab}$ is still negligible. We have to determine
if this region can occur when the complete vortex-vortex interaction is
taken into account or if it is a consequence of the local form of the
interaction used in the model. The complete Hamiltonian for interacting
vortex segments $n_\mu $ can be written in the generic form 
\begin{equation}
H=\sum_{i,j}\sum_{\mu =a,b,c}G_\mu \left( {\bf r}_i-{\bf r}_j\right) n_\mu
\left( {\bf r}_i\right) n_\mu \left( {\bf r}_j\right) .  \label{htotal}
\end{equation}
where $G_\mu \left( {\bf r}\right) $ is the non-local interaction function
between segments (in previous sections I used this type of Hamiltonian with
a local function $G$). The function $G_{ab}$ ($G_c$) is globally
proportional to an energy scale that we can identify with our $\varepsilon
_{ab}$ ($\varepsilon _c$) of equation (\ref{hamil}), and the naive
conclusion would be that in the highly anisotropic case ($\varepsilon
_{ab}\ll \varepsilon _c$) the system of interplane vortices has a
transition temperature $T_1$ of the order $k_BT_1$ $\simeq \varepsilon _{ab}
$, whereas the system of vortex-antivortex pairs within the planes has a
transition temperature $T_2$ of the order $k_BT_2$ $\simeq \varepsilon _c\gg
k_BT_1$. However, the structure of the function $G$ for segments parallel to
the layers 
$$
G_{ab}\left( {\bf k}\right) = 
$$
\begin{equation}
=\frac{\varepsilon _{ab}}{\left[ 4-2\cos\left%
(k_x\right)-2\cos\left(k_y\right)\right] +\frac{\varepsilon _{ab}%
}{\varepsilon _c}\left[ 2-2\cos\left(k_z\right) \right] +\frac{\varepsilon
_{ab}}\Delta }  \label{g12}
\end{equation}                
(where $\sqrt{\Delta /\varepsilon _{ab}}\equiv \lambda _c$ is the magnetic
penetration depth along the {\it c} direction in units of the lattice
spacing) shows that the range of the interaction of parallel to plane
segments increases with anisotropy as $\sim \left( \min \left[ \varepsilon
_c,\Delta \right] /\varepsilon _{ab}\right) ^{1/2},$ and this cannot be
neglected.\cite{korshunov,hetzel} In fact, in the language of the
renormalization group and in the case of very large $\lambda _c$,
interacting horizontal segments renormalize to non-interacting segments but
with an energy $\simeq \varepsilon _{ab}\left\{ \left( \varepsilon
_c/\varepsilon _{ab}\right) ^{1/2}\right\} ^2=\varepsilon _c$, so the
transition temperature $T_1$ is of the order of $k_BT_1\simeq \varepsilon _c.
$ A more detailed calculation\cite{korshunov} gives $T_1\sim 4\pi \min
\left[ \varepsilon _c,2\Delta \right] $, $T_2\sim \frac \pi 2\min \left[
\varepsilon _c,2\Delta \right] .$ To have finite $\rho _c$ at temperatures
where $\rho _{ab}$ is still zero is necessary at least that $T_1$ be lower
than $T_2,$ and this is not the case.

These estimates seem to rule out the practical possibility of a zone like
that for $\eta >\eta _2$ in figure \ref{eact}. However, the analysis made
corresponds to the case of zero external field and disorder, and we know
that the inclusion of them decreases the transition temperature of the
interplane system of loops. So the question if in a finite external field,
and for high anisotropies we can have a {\it c} axis transition temperature
lower than the corresponding to the {\it ab} plane has still to be answered.
The point cannot be fully discussed in all situations, but there is a
limiting case in which it can be addressed. In fact, the case with a
divergent anisotropy, in which the horizontal segments interact through a
really long range potential, can surprisingly be discussed much easily that
the case in which the interaction is local. Let us consider a finite value
of $\varepsilon _{ab}$, and take $\varepsilon _c,\Delta \rightarrow \infty ,
$ (so $T_2\rightarrow \infty $ also), and analyze the problem of the system
of horizontal vortex segments between two consecutive planes. To be able to
work with some indeterminations that will appear, the system size we will
consider the system size will be considered finite with a value $L\times L$, and the
limit $L\rightarrow \infty $ will be taken at the end. In this limiting case
the function $G_{ab}\left( {\bf k}\right) $ in terms of the two dimensional
vector ${\bf k\equiv }\left( k_x,k_y\right) ,$ reduces to $G_{ab}\left( 
{\bf k}\right) =\varepsilon _{ab}/\left[ 4-2\cos \left( k_x\right) -2\cos
\left( k_y\right) \right] .$ Introducing (integer) loop variables $l({\bf r})$ through the
substitution $\partial _\mu l\left( {\bf r}\right) =n_\mu \left( {\bf r}%
\right) ,$ and integrating by parts twice in the Hamiltonian, an equivalent
model is obtained: 
\begin{equation}
H=\varepsilon _{ab}\sum_{{\bf r}}\left( l\left( {\bf r}\right) -\bar{l}%
\right) ^2+I\sum_{{\bf r}}l\left( {\bf r}\right) ,  \label{hb}
\end{equation}
where $\bar{l}=\sum_{{\bf r}}l\left( {\bf r}\right) /L^2,$ is the mean value
of the loops variables over the $L^2$ plaquettes of the system. The time
derivative of $\bar{l}$ is proportional to the voltage generated by the
external current $I$. This expression is much easier to handle than the
original formulation, and it is a sort of mean field Hamiltonian in which
variables $l$ interact only through the $\bar{l}$ term. I stress however
that no approximations other than the ones mentioned were done on passing
from (\ref{htotal}) to (\ref{hb}). When $L$ is large, the partition function
corresponding to (\ref{hb}) can be calculated using a Lagrange multiplier for $%
\bar{l}$. The energy as a function of $\bar{l}$ becomes a periodic function
of $\bar{l}$ (except by the current term), and in the limit $T\gg
\varepsilon _{ab}$ it reads 
\begin{equation}
E=L^2\left( -\frac{T^2}{\varepsilon _{ab}}e^{-\pi ^2T/\varepsilon
_{ab}}\cos \left( 2\pi \bar{l}\right) +I\bar{l}\right) .  \label{edeb}
\end{equation}

When $L\rightarrow \infty ,$ and for any value of temperature there is a
critical value of $I,$ namely $I_{cr}=\frac{T^2}{\varepsilon _{ab}}e^{-\pi
^2T/\varepsilon _{ab}}$. This critical current is nonzero for any
temperature and thus indicates that the critical temperature of the system
is infinite. This is consistent with the exact results since we took $%
\varepsilon _c\rightarrow \infty $. Notice, however, that the extremely
small value of this critical current when $T$ is much larger that $%
\varepsilon _{ab}$ turns it difficult to verify this results in numerical
simulations.

When quenched disorder due to horizontal segments of externally generated
vortices are included, the previous picture changes in the following way.
The potential due to these quenched segments is also long ranged, and on the 
$l$ variables they produce a potential that goes as $\sim \varepsilon _{ab}/r$. Summing up
on each site the contributions from a random distribution of disorder on a
sample of size $L$ gives a potential of typical amplitude $V\sim \varepsilon _{ab}D\left[
\int^L\left( 1/r\right) ^2d^2r\right] ^{1/2}\sim \varepsilon _{ab}D\sqrt{2\pi \ln L}$, where $%
D$ is the concentration of quenched segments. The Hamiltonian now reads 
$$
H^D=\varepsilon _{ab}\sum_{{\bf r}}\left( l\left( {\bf r}\right) -\bar{l}%
\right) ^2+I\sum_{{\bf r}}l\left( {\bf r}\right) +
$$
\begin{equation}
+\varepsilon _{ab}D\sqrt{2\pi \ln L}\sum_{%
{\bf r}}\chi _{{\bf r}}l\left( {\bf r}\right) .  \label{hd}
\end{equation}                                    
The $\chi _{{\bf r}}$ are random variables with typical value $1$ and
satisfying $\sum_{{\bf r}}\chi _{{\bf r}}=0$.\cite{correl} In the same way
as before the energy of the system may be written in the form 
\begin{equation}
E^D=L^2\left( -\frac{T^2}{\varepsilon _{ab}}e^{-\pi ^2T/\varepsilon
_{ab}}e^{-\pi ^3D^2\ln L/2}\cos \left( 2\pi \bar{l}%
\right) +I\bar{l}\right),  \label{edebd}
\end{equation}
\begin{equation}
E^D=-\frac{T^2}{\varepsilon _{ab}}e^{-\pi ^2T/\varepsilon _{ab}}L^{2-\pi
^3D^2/2}\cos \left( 2\pi \bar{l}\right) +L^2I\bar{l}.
\end{equation}
When $L$ goes to infinity the first term will give an infinite activation
energy -and thus a zero resistivity- if $D<2/\pi
^{3/2}.$ On the other hand if $D>2/\pi ^{3/2}$ activation
energy goes to zero and the resistivity is finite. $D_{cr}=2
/\pi ^{3/2}$ is the critical value of the disorder in the system. As
we see, in this case the {\it c }axis transition occurs at zero temperature,
whereas the {\it ab} transition temperature is $\sim \varepsilon
_c\rightarrow \infty $.

Thus at least in the case of infinite ranged interactions between vortex
loops, it can be analytically shown that the {\it c} axis transition occurs
at lower temperature than the {\it ab }transition, if disorder is greater
than a critical value. This suggests that even in cases with finite $%
\varepsilon _c$, disorder can make longitudinal resistivity to appear at
lower temperatures than transversal dissipation. The
conclusion is that with the inclusion of the full interactions between
vortices and for finite values of $\varepsilon _{ab}$ and $\varepsilon _c$ (%
$\varepsilon _c\gg \varepsilon _{ab}$), a phase diagram as the one sketched
in Fig. \ref{ktargumento} still holds, but with the zero disorder
temperature transition being proportional to $\varepsilon _c$ rather to $%
\varepsilon _{ab}$. This indicates that a regime where $\rho_c\neq 0$ and 
$\rho_{ab}=0$ may be experimentally accessible in highly anisotropic 
and disordered samples.

\section{Summary and conclusion}

In this paper I have presented results from numerical simulations of vortex
lattices in the presence of external magnetic field and quenched disorder,
in which the interactions between vortices are neglected. In spite of the
simplicity of the model the results reproduce qualitatively well many
characteristics of the $\rho _c\left( T\right) $ and $\rho _{ab}\left(
T\right) $ functions observed in experiments, both for YBCO samples, in
which $\rho _c$ becomes different from zero when the value of $\rho _{ab}$
is already appreciable, and for BSCCO where experiments indicate that $\rho
_c\left( T\right) $ and $\rho _{ab}\left( T\right) $ have thermally
activated behaviors with the same activation energy. In addition, a new
regime in which activation energy for $\rho _c\left( T\right) $ is lower
than that for $\rho _{ab}\left( T\right) $ was found in the simulations,
corresponding to the case where coherence length along {\it c} direction
reduces to the distance between consecutive superconducting planes. 
Although in principle this regime
may be an artifact of having disregarded interactions, I have shown it may
occur in real samples in the presence of disorder, external magnetic
field, and for high enough anisotropy.

\section{Acknowledgments}

I want to thank C. A. Balseiro and M. Goffman for discussions, and K. Hallberg
for critical reading of the manuscript. This work was
financially supported by Consejo Nacional de Investigaciones
Cient\'{\i}ficas y T\'{e}cnicas (CONICET), Argentina.


\begin{references}
\bibitem{otranota}Resistivity should vanish discontinuously at $T_P$ in systems
of infinite size. This is not observed in Fig. \ref{r2d} due to finite size effects.
\bibitem{correl}  The variables $\chi _{{\bf r}}$ are spatially correlated,
however this is unimportant because the interaction between $l^{\prime }$s
is only through the mean field variable $\bar{l}$.
\end{references}
\end{document}